\documentstyle[12pt]{article}

\newcommand{\newc}{\newcommand}
\newc{\ra}{\rightarrow}
\newc{\lra}{\leftrightarrow}
\newc{\beq}{\begin{equation}}
\newc{\eeq}{\end{equation}}
\newc{\barr}{\begin{eqnarray}}
\newc{\earr}{\end{eqnarray}}

 1

\begin{document}
\begin{titlepage}
\begin{flushright}
{IOA.314/95}\\
\end{flushright}
\begin{center}
{\large \bf The role of nuclear form factors in Dark Matter
calculations}\\
\vspace{15mm} T. S. Kosmas  and  J. D. Vergados\\
\vspace{20mm}
Theoretical Physics Section, University of Ioannina,\\
 GR 451 10, Ioannina, Greece \\

\end{center}
\vspace{10mm}
\begin{abstract}
\vspace{8pt}

The momentum transfer dependence of the total cross section for elastic
scattering of cold dark matter candidates,
i.e. lightest supermymmetric particle (LSP), with nuclei is examined.
We find that even though the energy transfer is small ($\le 100 KeV$) the
momentum transfer can be quite big for large mass of the LSP and heavy nuclei.
The total cross section can in such instances be reduced by a factor of
about five.

\end{abstract}
\end{titlepage}
\newpage

There is ample evidence that about $90\%$ of the matter in the universe is
non-luminous and non-baryonic of unknown nature [1-3]. Furthermore, in order to
accommodate large scale structure of the universe one is forced to assume the
existence of two kinds of dark matter [3]. One kind is composed of particles
which were relativistic at the time of the structure formation. This is called
Hot Dark Matter (HDM). The other kind is composed of particles which were
non-relativistic at the time of stucture formation. These constitute the Cold
Dark Matter  (CDM) component of the universe. The COBE data [4] by examining
the inisotropy on background radiation suggest that the ratio of CDM to HDM is
2:1. Since about  $10\%$ of the matter of the universe is known to be baryonic,
we know that we have  $60\%$ CDM,  $30\%$ HDM and  $10\%$ baryonic matter.

The most natural candidates for HDM are the neutrinos provided that they have
a mass greater than $ 1 eV/c^2$. The situation is less clear in the case of
CDM.
The most appealing possibility, linked closely with Supersymmetry (SUSY), is
the
LSP i.e. the Lightest Supersymmetric Particle.

In recent years the phenomenological implications of Supersymmetry are
being taken very seriously [5-7]. Pretty accurate predictions at low energies
are now feasible in terms of few input parameters in the context of SUSY models
without any commitment to specific gauge groups. More or less such predictions
do not appear to depend on arbitrary choices of the relevant parameters or
untested assumptions.

In such theories derived from Supergravity the LSP is expected to be a neutral
fermion with mass in the $10-100 GeV/c^2$ region travelling with
non-relativistic
velocities ($\beta \simeq 10^{-3}$) i.e. with energies in the KeV region. In
the
absence of R-parity violation this particle is absolutely stable. But, even in
the presence of R-parity violation, it may live long enough to be a CDM
candidate.

The detection of the LSP, which is going to be denoted by $\chi_1$, is
extremely
difficult, since this particle interacts with matter extremely weakly. One
possibility is the detection of high energy neutrinos which are produced by
pair annihilation in the sun where this particle is trapped i.e. via the
reaction

\beq
\chi_1  + \chi_1 \ra \nu +{\bar \nu}
\label{eq:eg1}
\eeq

\noindent
The above reaction is possible since the LSP is a majorana particle, i.e. its
own antiparticle (\`{a} la $\pi^0$). Such high energy neutrinos can be
detected via neutrino telescopes.

The other possibility, to be examined in the present work, is the detection of
the  energy of the recoiling nucleus in the reaction

\beq
\chi_1  + (A,Z)  \ra \chi_1  + (A,Z)
\label{eq:eg2}
\eeq

\noindent
This energy can be converted into phonon energy and detected by a temperature
rise in cryostatic detector with sufficiently high Debye temperature [3,8,9].
The detector should be large enough to allow a sufficient number of counts but
not too large to permit anticoincidence shielding to reduce background.
A compromise of about $1Kg$ is achieved. Another possibility is the use of
superconducting granules suspended in a magnetic field. The heat produced
will destroy the superconductor and one can detect the resulting magnetix flux.
Again a target of about 1Kg is favored.

There are many targets which can be employed. The most popular ones contain the
nuclei
$^3_2He$, $^{19}_{9}F$, $^{23}_{11}Na$, $^{40}_{20}Ca$, $^{72,}$$^{76}_{32}Ge$,
$^{75}_{33}As$, $^{127}_{53}I$,
$^{134}_{54}Xe$, and $^{207}_{82}Pb$.

It has recently been shown that process (2) can be described by a four fermion
interaction [10-16] of the type [17]

\beq
L_{eff}  = -\frac{G_F}{\sqrt{2}} \, \big[ J_{\lambda} {\bar \chi_1}
\gamma^{\lambda}  \gamma^{5} \chi_1  \, +\,  J {\bar \chi_1}  \chi_1 \big]
\label{eq:eg3}
\eeq

\noindent
where

\beq
 J_{\lambda}  = {\bar N} \gamma_{\lambda} [\, f^0_V + f^1_V \tau_3
+( f^0_A + f^1_A \tau_3 ) \gamma_{5} \, ] N
\label{eq:eg4}
\eeq

\noindent
and

\beq
 J  = {\bar N} ( f^0_S + f^1_S \tau_3 ) N
\label{eq:eg5}
\eeq

\noindent
where we have neglected the uninteresting pseudoscalar and tensor
currents. Note that, due to the majorana nature of the LSP,
${\bar \chi_1} \gamma^{\lambda}  \chi_1 =0$ (identically).
The vector and axial vector form factors can arise out of Z-exchange and
s-quark
exchange [10-15] (s-quarks are the SUSY partners of quarks with spin zero).
They have uncertainties in them (see ref. [15] for three choices in the
allowed parameter space of ref. [5]).
The transition from the quark to the nucleon level is pretty straightforward in
this case. We will see later that, due to the majorana nature of the LSP, the
contribution of the vector current, which can lead to a coherent effect of all
nucleons, is suppressed [10-15]. Thus, the axial current, especially in the
case of light and intermediate mass nuclei, cannot be ignored.
The scalar form factors arise out of the Higgs exchange or via S-quark exchange
when there is mixing between s-quarks ${\tilde q}_L$ and ${\tilde q}_R$ [10-12]
(the partners of the left-handed and right-handed quarks). They have two types
of uncertainties in them [18]. One, which is the most important, at the quark
level due to the uncertainties in the Higgs sector. The other in going from
the quark to the nucleon level [16-17]. Such couplings are proportional to the
quark masses and hence sensitive to the small admixtures of $q {\bar q }$
(q other than u and d) present in the nucleon. Again values of $f^0_S$ and
$f^1_S$ in the allowed SUSY parameter space can be found in ref. [15].

The invariant amplitude in the case of non relativistic LSP takes the form
[15]

\barr
|{\it m}|^2 &=& \frac{E_f E_i -m_1^2 +{\bf p}_i\cdot {\bf p}_f } {m_1^2} \,
|J_0|^2 +  |{\bf J}|^2 +  |J|^2
 \nonumber \\ & \simeq & \beta ^2 |J_0|^2 + |{\bf J}|^2 + |J|^2
\label{eq:eg 45}
 \earr

\noindent
where $|J_0|$ and $|{\bf J }|$ indicate the matrix elements of the time
component and space component of the current $|J_\lambda|$ of eq. (4) and $J $
the matrix element of the scalar current J of eq. (5). Notice that $|J_0|^2$
is multiplied by $\beta^2$ (the suppression due to the majorana nature of LSP
mentioned above).
It is straightforward to show that

\beq
 |J_0|^2 = A^2 |F({\bf q}^2)|^2 \,\left(f^0_V -f^1_V \frac{N-Z}{A}
 \right)^2
\label{eq:eg7}
\eeq

\beq
 J^2 = A^2 |F({\bf q}^2)|^2 \,\left(f^0_S -f^1_S \frac{N-Z}{A}
 \right)^2
\label{eq:eg8}
\eeq

\beq
 |{\bf J}|^2 = \frac{1}{2J_i+1} |<J_i ||\, [ f^0_A {\bf \Omega}_0({\bf q})
\, + \, f^1_A {\bf \Omega}_1({\bf q}) ] \, ||J_i>|^2
\label{eq:eg 48}
\eeq
with

\beq
{\bf \Omega}_0({\bf q})  = \sum_{j=1}^A {\bf \sigma}(j) e^{-i{\bf q} \cdot
{\bf x}_j }, \qquad
{\bf \Omega}_1({\bf q})  = \sum_{j=1}^A {\bf \sigma} (j) {\bf \tau}_3 (j)
 e^{-i{\bf q} \cdot {\bf x}_j }
\label{eq:eg 49}
\eeq

\noindent
where ${\bf \sigma} (j)$, ${\bf \tau}_3 (j)$, ${\bf x}_j$ are the spin, third
component of isospin ($\tau_3 |p> = |p>$) and cordinate of the j-th nucleon and
$\bf q$ is the momentum transferred to the nucleus.

The differential cross section in the laboratory frame takes the form [15]

\barr
\frac{d\sigma}{d \Omega} &=& \frac{\sigma_0}{\pi} (\frac{m_1}{m_p})^2
\frac{1}{(1+\eta)^2} \xi  \{\beta^2 |J_0|^2  [1 - \frac{2\eta+1}{(1+\eta)^2}
\xi^2 ] + |{\bf J}|^2 + |J|^2 \}
\label{eq:eg 11}
 \earr

\noindent
where  $\eta = m_1/m_p A$  ( $m_p$ = proton mass), $\beta = \upsilon/c$
($\upsilon$ is the velocity of LSP), $m_1$ is the mass of LSP,
$\xi = {\bf {\hat p}}_i \cdot {\bf {\hat q}} \ge 0$ (forward scattering) and

\beq
\sigma_0  = \frac{1}{2\pi} (G_F m_p)^2 \simeq 0.77 \times 10^{-38}
cm^2 \label{eq:eg 12}
\eeq

\noindent
$|J_0|^2 $, $|{\bf J}|^2$ and $|J|^2$ are given by eqs. (7)-(9). The momentum
transfer  $\bf q$ is given by

\beq
|{\bf q}| = q_0 \xi, \qquad q_0 = \beta \frac{2  m_1 c }{1 +\eta}
\label{eq:eg 13}
\eeq

Some values of $q_0$ (forward momentum transfer) for some characteristic values
of $m_1$ and representative nuclear systems (light, intermediate and heavy)
are given in table 1. It is clear that the momentum transfer can be stable
for large $m_1$ and heavy nuclei.

The total cross section can be cast in the form

\barr
\sigma &=& \sigma_0 (\frac{m_1}{m_p})^2 \frac{1}{(1+\eta)^2} \,
 \{ A^2 \, [\beta^2 (f^0_V - f^1_A \frac{N-Z}{A})^2
\nonumber \\ & + &
(f^0_S - f^1_S \frac{N-Z}{A})^2 \,I_0(q^2_0) -
\frac{\beta^2}{2} \frac{2\eta +1}{(1+\eta)^2}
(f^0_V - f^1_V \frac{N-Z}{A})^2 I_1 (q^2_0) ]
\nonumber \\ & + &
(f^0_A \Omega_0(0))^2 I_{00}(q^2_0) - 2f^0_A f^1_A \Omega_0(0) \Omega_1(0)
I_{01}(q^2_0) \nonumber \\ & + &
(f^1_A \Omega_1(0))^2 I_{11}(q^2_0) \, \}
\label{eq:eg 14}
 \earr
where

\beq
I_\rho(q^2_0) = 2(\rho +1) \int_0^1 \xi^{1+2\rho} \, |F(q_0^2\xi^2)|^2 \,d\xi,
\qquad \rho = 0,1
\label{eq:eg 12}
\eeq

\beq
\Omega_\rho({\bf q}) = (2J_i+1)^{-\frac{1}{2}} < J_i||{\bf \Omega}_\rho({\bf
q})
||J_i>, \qquad \rho = 0,1
\label{eq:eg 12}
\eeq
(see eq. (10) for the definition of ${\bf \Omega}_\rho$) and

\beq
I_{\rho \rho^{\prime}}(q^2_0) = 2 \int_0^1 \xi \,
\frac{\Omega_\rho( q^2_0\xi^2)}{\Omega_\rho (0)} \,
\frac{\Omega_{\rho^{\prime}}( q^2_0\xi^2)}{\Omega_{\rho^{\prime}}(0)} \,d\xi
, \qquad \rho, \rho^{\prime} = 0,1
\label{eq:eg 12}
\eeq

In a previous paper [16] we have shown that the nuclear form factor can be
adequately  described within the harmonic oscillator model as follows

\beq
F({ q^2}) = \,[ \,
\frac{Z}{A} \Phi(qb,Z) + \frac{N}{A} \Phi(qb,N)\, ]\, e^{-q^2b^2/4}
\label{eq:eg 12}
\eeq
where $\Phi$ is a polynomial of the form [18]

\beq
\Phi(qb,\alpha)  = \sum_{\lambda =0}^{N_{max}(\alpha)}
\theta_\lambda^{(\alpha)}
(qb)^{2\lambda}, \qquad \alpha = Z, N
\label{eq:eg 12}
\eeq
$N_{max}(Z)$ and $N_{max}(N)$ depend on the major harmonic oscillator shell
occupied by protons and neutrons [16], respectively.
The integral $I_\rho(q^2_0)$ can be written as

\beq
I_\rho(q^2_0) \ra   I_\rho(u)  =
  \int_0^u x^{1+ \rho} \, |F( 2x/b^2)|^2 \,dx,
\label{eq:eg 12}
\eeq
where

\beq
u = q_0^2b^2/2, \qquad b=1.0 A^{1/3} \, fm
\label{eq:eg 15}
\eeq

With the use of eqs. (18), (19) we obtain

\beq
I_\rho(u)  = \frac{1}{A^2} \{ \, Z^2 I^{(\rho)}_{ZZ}(u) +
2NZI^{(\rho)}_{NZ}(u) +N^2I^{(\rho)}_{NN}(u) \}
\label{eq:eg 15}
\eeq
where

\beq
I^{(\rho)}_{\alpha \beta}(u)  =
\sum_{\lambda =0}^{N_{max}(\alpha)}  \, \sum_{\nu =0}^{N_{max}(\beta)}
\frac{\theta_\lambda^{(\alpha)}}{\alpha} \, \frac{\theta_\nu^{(\beta)}}{\beta}
\,
\frac{2^{\lambda +\nu+\rho} \,(\lambda +\nu+\rho)!}{u^{1+\rho}}
\Big[ 1 - e^{-u} \sum_{\kappa =0}^{\lambda +\nu+\rho} \,\,
\frac{u^\kappa}{\kappa!} \Big]
 \label{eq:eg 15}
\eeq
with $\alpha,\beta = N, Z$

 The coefficients  $\theta_\lambda^{(\alpha)}$ for light and medium nuclei have
been computed in ref. [16]. In table 2 we present them by including in addition
those for heavy nuclei.
The integrals $I_\rho(u)$ for three typical nuclei
( $^{40}_{20}Ca$, $^{72}_{32}Ge$  and $^{208}_{82}Pb$ ) are presented in fig. 1
as a function of $m_1$. We see that for light nuclei the modification
of the cross section by the inclusion of the form factor is small.
For heavy nuclei and massive $m_1$ the form factor has a dramatic effect on the
cross section and may decrease  it by a factor of about 5. The integral
$I_1(u)$
is even more suppressed but it is not very important.

The spin matrix elements depend on the details of the structure of the
nucleus considered. So is the spin form factor. The spin matrix element, since
it does not show coherence, is expected to be more important in the case
of odd light and intermediate nuclei. In the present work we will examine the
${\bf q}^2$ dependence of the spin matrix element in the cases of
$^{207}_{82}Pb$ and
$^{19}_{9}F$ whose structure is believed to be simple.

To a good approximation
[15,17] the ground state of the $^{207}_{82}Pb$
nucleus can be described as a $2s_{1/2}$
neutron hole in the  $^{208}_{82}Pb$  closed shell. One then finds

\beq
\Omega_0({\bf q}) \, = \,(1/\sqrt{3}) F_{2s} ({\bf q}^2), \qquad
\Omega_1({\bf q}) \, = \, -(1/\sqrt{3}) F_{2s} ({\bf q}^2)
\label{eq:eg 15}
\eeq
and

\beq
I_{00} = I_{01} = I_{11} = 2 \int_0^1 \xi \, [ F_{2s} (q^2) ]^2 \,d\xi
\label{eq:eg 15}
\eeq
Even though the probability of finding a pure $2s_{1/2}$ neutron hole
in the $\frac{1}{2}^-$ ground state of  $^{207}_{82}Pb$ is greater than 95\%,
the ground state magnetic moment is quenced due to the $1^+$ p-h excitation
involving the spin orbit partners. Hence we expect a similar suppression
of the isovector spin matrix elements. Thus we write

\barr
|{(1/2)}^->_{gs} \, & = & \, C_0 |(2s_{1/2})^{-1}> \Big[ \,
1 + C_1 |[0i_{11/2} (n) (0i_{13/2})^{-1} (n)] 1^+ >
\nonumber \\ & + &
C_2 |[0h_{9/2} (p) (0h_{11/2})^{-1} (p) ] 1^+ > + ... \Big]
\label{eq:eg 15}
\earr
Retaining terms  which are most linear in the coefficients $C_1$, $C_2$ we
obtain

\beq
\Omega_0({\bf q}) \, = \, C_0^2 \, \{ F_{2s} (q^2) /\sqrt{3}
 -8  \,[  (7/13)^{1/2} C_1 F_{0i} (q^2) \,
+\, (5/11)^{1/2} C_2 F_{0h} (q^2) ]\, \}
\label{eq:eg 15}
\eeq

\beq
\Omega_1({\bf q}) \, = \, C_0^2 \, \{ F_{2s} (q^2) /\sqrt{3}
 - 8  \,[  (7/13)^{1/2} C_1 F_{0i} (q^2) \,
-\, (5/11)^{1/2} C_2 F_{0h} (q^2) ]\, \}
\label{eq:eg 15}
\eeq
where

\beq
 F_{nl} (q^2) = e^{-q^2b^2/4} \sum_{\lambda =0}^{N_{max}}
\gamma_\lambda^{(nl)} (qb)^{2\lambda}
\label{eq:eg 15}
\eeq
The coefficients $\gamma_\lambda^{(nl)}$ are given in table 3.

The coefficients $C_0$,  $C_1$ and  $C_2$ were obtained by diagonalizing the
Kuo-Brown G-matrix [18,19] in a model space of 2h-1p configurations. Thus we
find

$$
C_0 = 0.973350, \qquad C_1 = 0.005295, \qquad  C_2 = -0.006984
$$
We also find

\beq
\Omega_0(0) \, = \,-(1/\sqrt{3}) (0.95659), \qquad (small \, \, \, retardation)
\label{eq:eg 15}
\eeq

\beq
\Omega_1(0) \, = \, -(1/\sqrt{3}) (0.83296), \qquad (sizable \, \, \,
retardation) \label{eq:eg 15}
\eeq

The amount of retardation of the total matrix element depends on the values  of
$f^0_A$ and $f^1_A$. Using eqs. (25) and (26) we can evaluate the integrals
$I_{00}$, $I_{01}$ and $I_{00}$. The results are presented in fig. 2. We see
that for a heavy nucleus and high LSP mass the momentum transfer dependence of
the spin matrix elements cannot be ignored.

In the second example we examine the spin matrix elements of the light nucleus
$^{19}_9F$. Assuming that the ground state wave function is a pure $SU(3)$
state with the largest symmetry i.e $f=[3], (\lambda\mu) = (60)$, we obtain
[20,21] the expression

\beq
\frac {\Omega_1({\bf q})}{\Omega_1(0)} \, = \, \frac{4}{9} \,  F_{2s} (q^2)
 +  \frac{5}{9} F_{0d} (q^2)
\label{eq:eg 15}
\eeq

There is only isovector component contribution (the isoscalar matrix
element vanishes). The results for the
$I_{11}(u)$ integral are shown in fig. 3.
We see that the effect of the nuclear form factor on the cross section
for light nuclei is insignificant.

In the present paper we have examined the momentum transfer dependence of the
nuclear matrix elements entering the elastic scattering of cold dark matter
candidates (LSP) with nuclei. We have found that such a momentum transfer
dependence is very pronounced for heavy nuclear targets and mass of the LSP
in the 100 GeV region.

\newpage

\vspace{0.5cm}


\newpage

{\bf Figure Captions}
\vskip0.3cm

Fig. 1. The integral $I_0(u)$, which describes the  main coherent
contribution to the total cross section as a function of the LSP mass
($m_1$), for three typical nuclei: $^{40}_{20}Ca$, $^{72}_{32}Ge$ and
$^{208}_{82}Pb$.

\vskip0.2cm

Fig. 2. The integral $I_1(u)$, entering
the total coherent cross section as a function of the LSP mass ($m_1$),
for three
typical nuclei: $^{40}_{20}Ca$, $^{72}_{32}Ge$ and $^{208}_{82}Pb$. For its
definition see eqs (11) and (15) of the text.

\vskip0.2cm

Fig. 3. The integral $I_{11}$,associated with the
spin isovector - isovector matrix elements for  $^{207}_{82}Pb$ and
$^{19}_{9}F$
as a function of the LSP mass ($m_1$). The other two intergals $I_{00}$ and
$I_{01}$  are almost identical and are not shown.

\newpage
\begin{table}
{\bf Table 1} : The quantity $q_0$ (forward momentum transfer) in units of
$fm^{-1}$ for three values of $m_1$ and three typical nuclei.

\vskip0.4cm

\begin{tabular}{|c|ccc|}
\hline
Nucleus &   $m_1=30 GeV$    &   $m_1=100 GeV$ &   $m_1=150 GeV$ \\
\hline
     $^{40}_{20}Ca$   &  .174   &  .290 &  .321 \\
     $^{72}_{32}Ge$   &  .215   &  .425 &  .494 \\
     $^{208}_{82}Pb$  &  .267   &  .685 &  .885 \\
\hline
\end{tabular}
\end{table}


\vskip2.0cm

\begin{table}
{\bf Table 2.}
The coefficients $\theta_{\lambda}$ determining the proton
and neutron form factors for all closed (sub)shell nuclei.
In a harmonic oscillator basis they are rational numbers.
The coeficients for $\lambda=0$ are equal to $Z$ (or $N$).

\vskip0.4cm

\begin{tabular}{|c|rllllll|}
\hline
  nlj-level &  $\,\,\,\lambda =0\,\,\,$ &  $\lambda =1$ &
$\lambda =2$ &  $\lambda =3$ &  $\lambda =4$
&  $\lambda =5$ &  $\lambda =6$ \\
\hline
  $0s_{1/2}$ &  2 &  &  &  &  &  &\\
  $0p_{3/2}$ &  6 & -2/3 &  &  &  & & \\
  $0p_{1/2}$ &  8 & -1 &  &  &  &  &\\
  $0d_{5/2}$ &  14 & -3 & 1/10 &  &   & & \\
  $1s_{1/2}$ &  16 & -11/3 & 11/60 &  &  &  &\\
  $0d_{3/2}$ & 20 & -5 & 1/4 &  &  &  &\\
  $0f_{7/2}$ &  28 & -9 & 13/20 & -1/105 &  &  &\\
  $1p_{3/2}$ & 32 & -11 & 61/60 & -11/420 &  & & \\
  $0f_{5/2}$ &  38 & -14 & 79/60 & -1/30 &  &  &\\
  $1p_{1/2}$ &  40 & -15 & 3/2 & -1/24 &  &  &\\
  $0g_{9/2}$ &  50 & -65/3 & 5/2 & -5/56 & 1/1512 &  &\\
  $0g_{7/2}$ &  58 &  -27 & 33/10  & -107/840 & 1/840 & &\\
  $1d_{5/2}$ &  64 & -31 & 17/4  & -173/840  & 1/336 & &\\
  $1d_{3/2}$ &  68 & -101/3 & 293/60 & -31/120 & 1/240 & & \\
  $2s_{1/2}$ &  70 & -35  & 21/4 & -7/24 & 1/192  & &\\
  $0h_{11/2}$ &  82 & -45 & 29/4 & -73/168 & 37/4032 & -1/27720 &\\
  $0h_{9/2}$ &  92 & -160/3 & 107/12 & -31/56 & 151/12096 & -1/15120& \\
  $1f_{7/2}$ & 100 & -60 & 217/20 & -653/840 &449/20160 & -1/5040 & \\
  $1f_{5/2}$ & 106 & -65 & 123/10 & -397/420 &199/6720 & -1/3360 & \\
  $2p_{3/2}$ & 110 & -205/3 & 403/30 & -153/140 &253/6720 &-1/2240 & \\
  $2p_{1/2}$ & 112 & -70 & 14 & -7/6 &1/24 &-1/1920 & \\
  $0i_{13/2}$ & 126 & -84 & 35/2 & -3/2 &1/18 &-49/63360 &1/617760 \\
\hline
\end{tabular}
\end{table}

\newpage
\begin{table}
{\bf Table 3.}
The coefficients $\gamma_{\lambda}^{(nl)}$, entering the polynomial
describing the form factor (see eq. (29)) of a single particle harmonic
oscillator wave function up to  $6 \hbar \omega$, i.e. throughout the
 periodic table.

\vskip0.4cm

\begin{tabular}{|c|cllllll|}
\hline
 $n$ $l$ &  $\,\,\,\lambda =0\,\,\,$ &  $\lambda =1$ &
$\lambda =2$ &  $\lambda =3$ &  $\lambda =4$
&  $\lambda =5$ &  $\lambda =6$ \\
\hline
 0 0 &  1 &  &  &  &  &  &\\
 0 1 &  1 & -1/6 &  &  &  & & \\
 1 0 &  1 & -1/3 & 1/24   &  &  &  &\\
 0 2 &  1 & -1/3 & 1/60   &  &   & & \\
 1 1 &  1 & -1/2 & 11/120 &-1/240  &  &  &\\
 0 3 &  1 & -1/2 & 1/20   &-1/840  &  &  &\\
 2 0 &  1 & -2/3 & 11/60  &-1/60   & 1/1920  &  &\\
 1 2 &  1 & -2/3 & 19/120 &-11/840 & 1/3360  &  & \\
 0 4 &  1 & -2/3 & 1/10   &-1/210  & 1/15120 &  &\\
 2 1 &  1 & -5/6 & 17/60  &-31/840 & 9/4480  &-1/26880 &\\
 1 3 &  1 & -5/6 & 29/120 &-47/1680& 37/30240&-1/60480 &\\
 0 5 &  1 & -5/6 & 1/6    &-1/84   & 1/3024  &-1/332640&\\
 3 0 &  1 &  -1  & 17/40  &-31/420 & 27/4480 &-1/4480  & 1/322560 \\
 2 2 &  1 &  -1  & 2/5    &-1/15   & 41/8064 &-1/5760  & 1/483840 \\
 1 4 &  1 &  -1  & 41/120 &-1/20   & 1/315   &-1/11880 & 1/1330560 \\
 0 6 &  1 &  -1  & 1/4    &-1/42   & 1/1008  &-1/55440 & 1/8648640 \\
\hline
\end{tabular}
\end{table}

\end{document}